\documentclass[twocolumn,prl,runinaddress,showpacs]{revtex4}
\usepackage{amssymb}
\usepackage{graphicx}

\input{tcilatex}

\begin{document}

\title{Edge Spin Current and Spin Polarization in Quantum Hall Regime}
\author{Yun-Juan Bao, Huai-Bing Zhuang, Shun-Qing Shen, and Fu-Chun Zhang*}
\affiliation{Department of Physics, and Center for Theoretical and Computational Physics,
The University of Hong Kong, Pokfulam Road, Hong Kong, China}
\date{May 5, 2005}

\begin{abstract}
We study the edge spin current and spin polarization and their repsonses to
an external electric field in a two-dimensional electron gas with a Rashba
spin-orbit coupling in the quantum Hall regime. The edge state carries a
large spin current, which drops sharply away from the boundary. The spin
Hall current of the edge states has a resonance at a critical magnetic
field, accompanying a spin rotation in the bulk. The spin Hall current is
shown to be proportional to the spin polarizationin, which provides an
explicit way to extract the spin current in experiment.
\end{abstract}

\maketitle

The edge state and edge charge current in two-dimensional electron gas
(2DEG) in a magnetic field have been well studied in the past two decades
and have played an important role in understanding the quantum Hall effect 
\cite{Laughlin81prb,Halperin82prb,MacDonald84prb,Wen92}. Recent progresses
on spintronics stimulate extensive study of spin generation and transport in
semiconductors \cite{Prinz98Science}. The relativistic quantum effect of
moving electrons in a confining potential in 2DEG induces a spin-orbit
coupling to split the energy spectra of electrons, and provides an efficient
way to control the electron's spin. Driven by an external electric field
moving electrons may generate a pure spin Hall current via the magnetic
impurity or spin-orbit coupling \cite{Dyakonov71}. Very recently the spin
accumulation has been observed in both n-type semiconductors and p-type
heterojunctions in experiments, which provides substantial evidence of pure
spin current \cite{Kato04Science}. In the presence of a strong magnetic
field, the competition between the Rashba spin-orbit coupling and the Zeeman
splitting in 2DEG introduces an additional degeneracy and gives arise to
some novel effects, such as the resonant spin Hall conductance \cite%
{Shen04prl} and the anisotropic spin transport \cite{Schliemann03prb}. In
this Letter we study the edge spin current and spin polarization, and their
responses to an external electric field in 2DEG with a Rashba coupling in a
magnetic field. The edge effect will cause the anticrossing of Landau
levels. The spin Hall conductance is calculated using the edge state
approach, and the result can be used to complement the bulk theory for the
resonant spin Hall effect. The spin current is shown to be proportional to
the spin polarization along the $y$-direction in an finite magnetic field,
which provides an explicit way to extract the spin current in experiments.

We consider a 2DEG with the Rashba coupling in the x-y plane of area $%
L\times L$ subject to a perpendicular magnetic field $\vec{B}=-B\hat{z}$.
The electrons are confined between $-L/2$ and $L/2$ in the $y$-direction by
an infinite potential wall, and its wavefunction is periodic along the $x$%
-direction. We choose a Landau gauge $\vec{A}=By\hat{x}$. The Hamiltonian
for a single electron with a Rashba coupling is given by 
\begin{equation}
H=\vec{\Pi}^{2}/2m-g_{s}\mu _{B}B\sigma _{z}/2+(\lambda /\hbar )(\Pi
_{x}\sigma _{y}-\Pi _{y}\sigma _{x}),
\end{equation}%
where the confining potential is implied. $m$, $-e$, and $g_{s}$ are the
electron's effective mass, charge, and Lande-g factor, respectively. And $%
\vec{\Pi}=\vec{p}+e\vec{A}/c$ is the kinetic operator, $\mu _{B}$ is the
Bohr magneton, and $\sigma _{\alpha }$ are the Pauli matrices. The edge
state and the edge charge current in the absence of the Rashba coupling have
been studied previously \cite{Halperin82prb,MacDonald84prb}. In that case,
the eigen state is given by $\Phi _{n,y_{0},s}=e^{iy_{0}x/l_{b}^{2}}\phi
_{n,y_{0}}(y)\chi _{s}/\sqrt{2\pi }$, with $n$ the Landau level index, $%
p_{x}=$ $\hbar y_{0}/l_{b}^{2}$ the $x$-component momentum quantum number or
orbital center of the wave function $\phi _{n,y_{0}}(y),$ and the magnetic
length $l_{b}=\sqrt{\hbar c/eB}$. $\chi _{\pm 1/2}$ is the eigenstate of
spin $S_{z}=\pm 1/2$. The wave function $\phi _{n,y_{0}}(y)$ is the
confluent hypergeometric function determined by the eigen value equation $%
a_{y_{0}}^{\dag }a_{y_{0}}\phi _{n,y_{0}}=\nu _{n,y_{0}}\phi _{n,y_{0}}$,
satisfying the boundary condition $\phi _{n,y_{0}}(\pm L/2)=0$, where $%
a_{y_{0}}=\left( y+y_{0}+i\frac{c}{eB}p_{y}\right) /\sqrt{2}l_{b}$ and $%
[a_{y_{0}},a_{y_{0}^{\prime }}^{\dag }]=\delta _{y_{0},y_{0}^{\prime }}$.
The energy eigenvalues are $E_{n,s}^{(0)}(y_{0})=\left[ \nu
_{n,y_{0}}+(1-g_{s})/2\right] \hbar \omega $ where $\omega =eB/mc$ is the
cyclotron frequency. For $\left\vert y_{0}\right\vert \ll L/2$ far away from
the edges, the problem is reduced to a simple harmonic oscillator and the
eigen value $\nu _{n,y_{0}}=n$, a non-negative integer. At the two edges $%
y_{0}=\pm L/2$, the boundary condition leads to $\nu _{n,y_{0}}=2n+1$.

In the presence of the Rashba spin-orbit coupling, $p_{x}$ remains to be a
good quantum number. In the Hilbert subspace of $y_{0}$ the Hamiltonian can
be written as 
\begin{equation}
H(y_{0})=\hbar \omega \left[ a_{y_{0}}^{\dag }a_{y_{0}}+(1-g\sigma
_{z})/2+i\eta (a_{y_{0}}\sigma _{-}-a_{y_{0}}^{\dag }\sigma _{+})\right] ,
\label{coupling}
\end{equation}%
where $\eta =\lambda ml_{b}/\hbar ^{2}$ is the effective Rashba coupling,
and $g=g_{s}m/2m_{e}$. Far away from the two edges the Rashba coupling mixes
only two states $\Phi _{n,y_{0},-1}$ and $\Phi _{n+1,y_{0},1}$ so that an
analytical solution can be obtained \cite{Shen04prl,Schliemann03prb,Rashba60}%
. Near the boundary, exact analytical solutions seem unlikely. We use $%
\left\{ \Phi _{n,y_{0},1},\Phi _{n,y_{0},-1}\right\} $, the eigen states of $%
H(y_{0})$ at $\lambda =0,$ as the base functions. The lower energy spectra $%
\{E_{n,s}(y_{0})\}$ and the corresponding eigenvectors can be obtained
numerically by truncating sufficiently higher Landau level states. In Fig.
1, we show a typical energy spectrum of the lowest twenty Landau levels at $%
y_{0}=L/2-4l_{b}$ as a function of the effective Rashba coupling $\eta $,
for a set of realistic parameters suitable for In$_{053}$Ga$_{0.47}$As/In$%
_{0.52}$Al$_{0.48}$As \cite{Nitta97prl}. While the quantitative results
depend on the precise parameters, the basic physics we address is quite
general. Different from the bulk state, where the competition between the
Rashba coupling and the Zeeman splitting introduces an additional degeneracy 
\cite{Shen04prl}, there is a level anticrossing \cite{Reynoso04prb} near the
boundary of a distance of several magnetic length as shown in Fig. 1. The
edge effect to the level anticrossing may be understood as follows. The
Rashba coupling mixes the two edge states of opposite spins in the same
Landau level, namely $\Phi _{n,y_{0},-1}$ and $\Phi _{n,y_{0},1}$, while
this mixing vanishes for the bulk states.

\begin{figure}[tbp]
\includegraphics*[width=8.5cm]{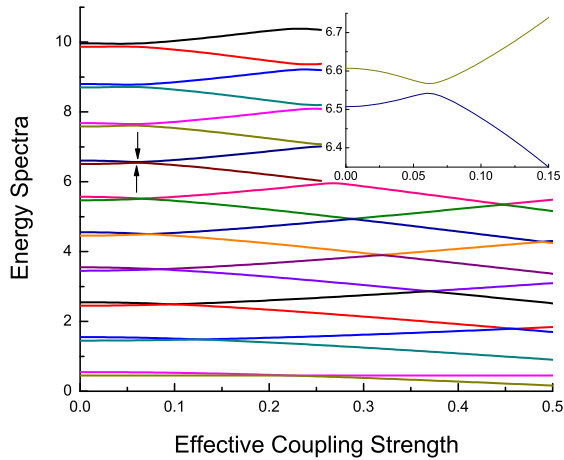}
\caption{ Energy spectra in the unit of $\hbar \protect\omega $ versus the
effective Rashba coupling $\protect\eta =\protect\lambda ml_{b}/\hbar ^{2}$
at the edge $y_{0}=\pm (L/2-4l_{b})$. The parameters are $\protect\lambda %
=0.96\times 10^{-4}\hbar $c$,$ $n_{e}=1.9\times 10^{12}/$cm$^{2},g_{s}=4,$
and $m=0.05m_{e}$, suitable for In$_{053}$Ga$_{0.47}$As/In $_{0.52}$Al$
_{0.48}$ As in Ref ~\protect\cite{Nitta97prl}. The inset figure shows the
level anticrossing of the edge state. }
\end{figure}

We now turn to the discussion of the charge and spin currents and the spin
polarization near the boundary. The charge current operator of a single
electron, in the Hilbert subspace of $y_{0}$, is given by, $j_{c}=-ev_{x}$,
and $v_{x}=[x,H]/i\hbar $. The spin-$\alpha $ component current operator is $%
j_{s}^{\alpha }=\hbar (\sigma _{\alpha }v_{x}+v_{x}\sigma _{\alpha })/4$.
Let $(j_{c,s})_{\tau }$ be the charge (c) or spin (s) current carried by an
electron in the eigen state $\left\vert \tau \right\rangle $ of $H$ in
Eq.(2) with a collective index $\tau =(n,y_{0},s)$ and $\left( S^{\alpha
}\right) _{\tau }=\left\langle \tau \right\vert \sigma _{\alpha }\left\vert
\tau \right\rangle \hbar /2$ the average $\alpha $-component spin
polarization in that state. Note that the state represented by $y_{0}$ is
extended in the $x$-direction but localized around $y=-y_{0}$ in the $y$%
-direction. The calculated expectation values are plotted in Fig. 2 as
functions of the position of the edge state described by $y_{0}=L/2-\zeta
l_{b}$ $(\zeta \geq 0).$ In comparison with their bulk values, both $%
(j_{c})_{\tau }$ and $(j_{s}^{z})_{\tau }$ are markedly larger. As for the
spin polarization, $\left( S^{x}\right) _{\tau }=0$, while $\left(
S^{y}\right) _{\tau }$ and $\left( S^{z}\right) _{\tau }$ display
interesting features: the spin is mostly polarized along the $z$-axis in the
bulk, while mostly polarized along the $y$-axis at the edges. It is
interesting to note from Fig. 2 that the large polarization of $S_{y}$
coincides with the large charge current at the edge. From a mean field point
of view, the Rashba coupling $V_{R}(y_{0})\equiv $ $\lambda m(v_{x}\sigma
_{y}-v_{y}\sigma _{x})/\hbar -2m\lambda ^{2}/\hbar ^{2}$ introduces an
effective magnetic field along the $y$-direction, which is proportional to
the charge current along the $x-$direction. It is most interesting to note
that the average of $S_{y}$ is identical to the spin current $j_{s}^{z}$
except for an overall proportionality, $\left( j_{s}^{z}\right) _{\tau
}=-0.831(\hbar /ml_{b})(S^{y})_{\tau }$ for the specific parameters in Fig.
2. In fact, there is an exact relation between the two quantities in the
present case, 
\begin{equation}
\lambda \left( j_{s}^{z}\right) _{\tau }=-\left( g_{s}\hbar \mu
_{B}/2m\right) B(S^{y})_{\tau }.  \label{relation}
\end{equation}%
To derive the above relation and to see the required condition for the
equation, we consider a commutator \cite{Dimitrova04xxx}, 
\begin{equation}
\lbrack H,\sigma _{x}]=-i\left( 4m\lambda /\hbar ^{2}\right)
j_{s}^{z}-ig_{s}\mu _{B}B\sigma _{y}.
\end{equation}
In the present case, $p_{x}$ is a good quantum number, and the energy
eigenstate $\left\vert \tau \right\rangle $ is localized in the $y$%
-direction around $-y_{0}$ and vanishes at the two edges. Because of these
properties, the expectation value of this commutator $\left( [H,\sigma
_{x}]\right) _{\tau }=\left\langle \tau \right\vert H\sigma _{x}-\sigma
_{x}H\left\vert \tau \right\rangle =0,$ and hence Eq. \ref{realtion} is
approved. We note that the relation holds in the presence of an electric
field $\vec{E}=E\hat{y}$. Because the relation applies to each eigenstate of 
$H$ in the absence or in the presence of an electric field, the
thermodynamic averages of the spin polarization and the spin current are
also proportional. However, one has to be cautious to apply the relation to
the case with $B=0$ and $E\neq 0$. At $B=0$, the states are extended in both 
$x$- and $y$-directions, and the above derivation for the vanishing integral
may not apply. We note that the spin Hall current at $B=0$ cannot be
obtained by taking $B\rightarrow 0$. Inclusion of the Dresselhaus coupling 
\cite{Shen04prb} will also give a relevant relation. In the presence of
disorder and electron-electron interactions, the spin current and spin
polarization vary in space, but such a relation between the spin current and
the spin polarization can be shown to hold at any spatial point in the
quantum Hall regime. The detailed discussion will be presented elsewhere.

\begin{figure}[tbp]
\includegraphics*[width=8.5cm]{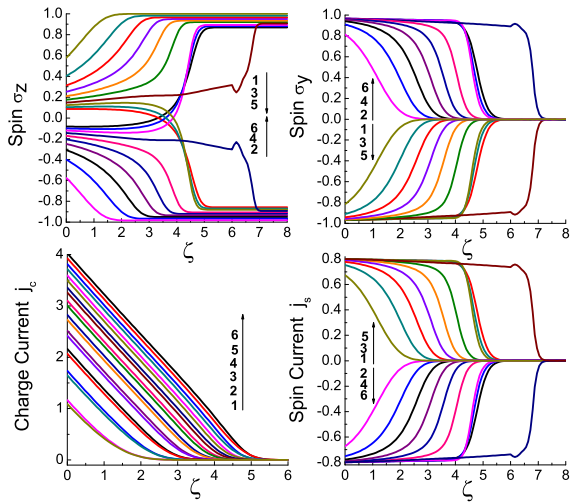}
\caption{ The charge Hall current $\ $in unit of $-el_{b}\protect\omega $ ,
the spin Hall current in unit of $\hbar l_{b}\protect\omega /2$, $S^{z}$ and 
$S^{y}$ in unit of $\hbar /2$ in the twenty lowest energy states near the
boundary $y_{0}=L/2-\protect\zeta l_{b}.$ The charge and spin currents as
well as $S^{y}$ are antisymmetric, while $S^{z}$ is symmetric with respect
to $y_{0}.$ The magnetic field $B=6.1$T, for which the 13$^{rd}$ and 14$^{th}
$ levels are degnerated far away from the two edges. \protect\cite{Shen04prl}
The arrowed line with numbers indicate the order of the energy eigenstates
from the lowest to higher ones.}
\end{figure}

Now we come to discuss the effect of an electric field along the $y$%
-direction, $V_{E}=-eEy$. We use the truncation approximation to solve the
problem and evaluate the energy eigenvalues and eigenstates numerically. We
find that the edge charge current responses linearly to the electric field.
Here we focus on the non-linear behaviors of the spin currents and spin
polarization at or near the resonant magnetic field $B_{0}$ where the two
levels in the bulk are degenerate \cite{Shen04prl}. In Fig. 3, we plot $%
\left( j_{s}\right) _{\tau }$, $\left( S^{y}\right) _{\tau }$ and $\left(
S^{z}\right) _{\tau }$ in a weak electric field $E=0.01$volt/m. Our data
show that $\left( j_{s}\right) _{\tau }$ is still proportional to $\left(
S^{y}\right) _{\tau }$ in the presence of the electric field so we plot $%
\left( j_{s}\right) _{\tau }$ and $\left( S^{y}\right) _{\tau }$ in the same
figure. At the resonant point we notice that the electric field generates a
finite spin current in the bulk which is almost equal to the edge spin
current at the edge of $y_{0}=-L/2.$ At the edge of $y_{0}=+L/2$ the current
has the same value, but different direction. $\left( S^{z}\right) _{\tau }$
decreases quickly to zero, but the $\left( S^{y}\right) _{\tau }$ increases
to one in unit of $\hbar /2$ approximately. Thus the spin rotates from the $z
$- to $y$-direction. Near the resonant magnetic field where the energy gap $%
E_{G}$ of the two levels at the bulk is larger than, but comparable with the
electric field energy, $E_{G}\approx eEl_{b},$ both the spin current and
spin polarization vary non-linearly with the electric field. When the energy
gap $E_{G}$ is much larger than the electric field energy, $E_{G}\gg eEl_{b}$
, the response becomes linear to the field. The rotation of the spins of the
two levels in a weak field can be understood in the theory of resonant spin
Hall effect \cite{Shen04prl}. A weak field removes the degeneracy of the two
levels of the bulk along the spin $z$-direction, and as a result the mixed
two levels give the spin along the $y$-direction. The rotation of spin from
the spin $z$-direction to the spin $y$-direction should be observed near the
resonant point experimentally.

\begin{figure}[tbp]
\includegraphics*[width=8.5cm]{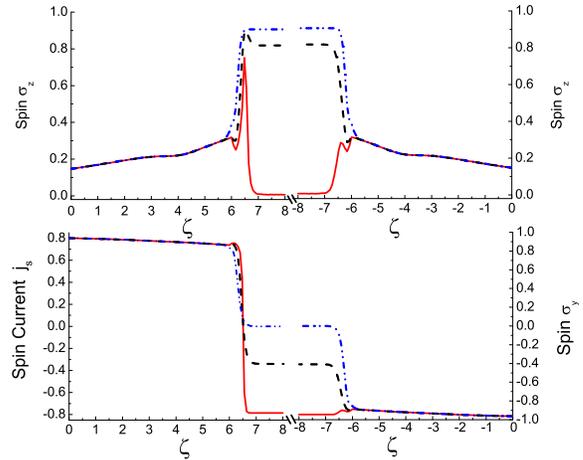}
\caption{ The spin polarization and spin current of the two levels which are
almost degenerated subjected to an fixed external electric field, $E=0.01$
volt/m. The left is for $y_{0}=L/2-\protect\zeta l_{b}$ and the right are
for $y_{0}=-L/2-\protect\zeta l_{b}$. (a) The solid line is the two levels
with an energy gap $\Delta E$ in the bulk less than the electric field
energy $eEl_{b}$; (b) the dashed line for the two levels with $\Delta
E\simeq eEl_{b}$; (c) the dashed-dotted line for the two levels of $\Delta
E\gg eEl_{b}$. }
\end{figure}

With the periodic boundary condition along the $x$-direction the velocity
operator in the Hilbert subspace of $y_{0}$ can be written as $%
v_{x}=(l_{b}^{2}/\hbar )(\partial H/\partial y_{0})$ and thus the spin
current operator $j_{s}^{z}=l_{b}^{2}(\partial H/\partial y_{0}\sigma
_{z}+\sigma _{z}\partial H/\partial y_{0})/2.$ Following the works by
Halperin \cite{Halperin82prb} and MacDonald and Streda \cite{MacDonald84prb}
the\ total spin Hall current in the filled Landau level ($n,s$) can be
expressed as 
\begin{eqnarray}
\left( j_{s}^{z}\right) _{n,s} &=&\int_{-L/2}^{+L/2}\frac{dy_{0}}{4\pi }%
\frac{\partial E_{n,s}(y_{0})}{\partial y_{0}}\left\langle \tau \right\vert
\sigma _{z}\left\vert \tau \right\rangle -  \nonumber \\
&&\sum_{n^{\prime },s^{\prime }}\int_{-L/2}^{+L/2}\frac{dy_{0}}{8\pi }%
(E_{n,s}(y_{0})-E_{n^{\prime },s^{\prime }}(y_{0}))\times   \nonumber \\
&&\left( \left\langle \tau \right\vert \partial _{y_{0}}\left\vert \tau
^{\prime }\right\rangle \left\langle \tau ^{\prime }\right\vert \sigma
_{z}\left\vert \tau \right\rangle -\left\langle \tau \right\vert \sigma
_{z}\left\vert \tau ^{\prime }\right\rangle \left\langle \tau ^{\prime
}\right\vert \partial _{y_{0}}\left\vert \tau \right\rangle \right) 
\end{eqnarray}%
where $\tau ^{\prime }=\left( n^{\prime },y_{0},s^{\prime }\right) .$
Without the spin-orbit coupling the energy eigenstate satisfies $%
\left\langle \tau \right\vert \sigma _{z}\left\vert \tau ^{\prime
}\right\rangle =s\delta _{n,n^{\prime }}\delta _{s,s^{\prime }}$ so that $%
\left( j_{s}^{z}\right) _{n,s}=-seV/4\pi ,$ which is only determined by the
voltage difference at the two edges, $E_{n,s}(L/2)-E_{n,s}(-L/2)=-eV$, and
the inclusion of impurities and Coulomb interactions in the Hamiltonian does
not affect this result as for the charge Hall current \cite%
{Laughlin81prb,Halperin82prb}. The spin Hall conductance displays a series
of plateau in the quantum Hall regime, $G_{s}=(1-(-1)^{n})e/8\pi $
corresponding to the quantum Hall conductance, $G_{c}=ne^{2}/h.$ In the
presence of spin-orbit coupling the states with different spins will be
mixed together, and the spin gradually deviates from the $z$- to $y$%
-direction and the spin Hall conductance varies with the effective Rashba
coupling or magnetic field through tuning the energy gap between the two
states especially near the Fermi level. We calculate the total spin Hall
conductance numerically as a function of $1/B$ for a fixed chemical
potential \textit{\ assuming that the voltage drops only near the edges and
the bulk state does not contribute to the total spin Hall current}. Both
charge and spin Hall conductances are plotted in Fig. 4. As expected the
Hall conductance is quantized, the spin Hall conductance has the order of $%
e/4\pi $ if odd number of Landau levels is occupied, and is of order of $%
10^{-3}$ -- $10_{{}}^{-4}e/4\pi $ if even number of Landau levels is
occupied. The spin Hall conductance is a function of the effective
spin-orbit coupling, which varies with the magnetic field. The resonant peak
appears only when the two degenerate bulk Landau levels crosses over a
special value of chemical potential with decreasing the magnetic field. In
the inset of Fig. 4 we extract the spin Hall conductance for a fixed density
of charge carriers from the results for a fixed chemical potential if $L\gg
l_{b}$. The values of the spin Hall conductance are consistent with the bulk
theory for the fully filled Landau levels \cite{Shen04prl}.

\begin{figure}[tbp]
\includegraphics*[width=8.5cm]{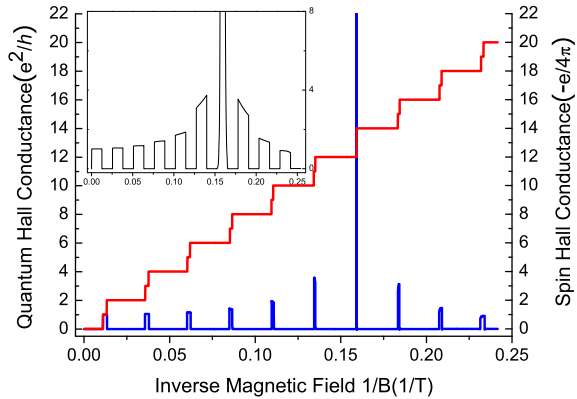}
\caption{ The spin and charge Hall conductance as a function of $1/B$ for a
fixed chemical potential, $\protect\mu =14.539$meV$.$ The inset is spin Hall
conductance for a fixed density of charge carriers, $n_{e}=1.9\times 10^{12}$
cm$^{-2}$. }
\end{figure}

The spin and charge edge currents will become dominant when the bulk states
are localized due to the impurities and the charge Hall conductance is
quantized. In a realistic sample of a finite size both charge and spin bulk
Hall current densities will not be suppressed completely, but decays with
the size of the sample if the impurities are taken into account. The total
charge and spin Hall currents consist of both the edge and bulk current. The
ratio of the edge and bulk Hall currents or the distribution of the electric
field are determined by the disorder, Coulomb interaction, and the rigidity
of the confining potential wall as studied by several authors \cite%
{Thouless93prl}. The quantum Hall effect can be understood from both the
bulk and edge state point of view. In this Letter we have presented a
picture of edge states for spin Hall conductance for a 2DEG with a Rashba
coupling. The theory is consistent with and complementary to the bulk theory
for resonant spin Hall conductance, especially in quantum Hall regime. In
this system the motion of electrons will induce an effective magnetic field
via the Rashba coupling, and generate spin polarization. Since the spin
current is proportional to the spin polarization along the $y$-direction
this relation will provide an explicit way to extract spin current from the
distribution of spin polarization.

The authors would like to thank Kun Yang and Qian Niu for helpful
discussions. This work was supported by the Research Grant Council of Hong
Kong.


\begin{thebibliography}{*}
\bibitem[*]{leave} On leave from Department of Physics, University of
Cincinnati, Ohio

\bibitem{Laughlin81prb} R. B. Laughlin, Phys. Rev. B 23, 5632 (1981).

\bibitem{Halperin82prb} B. I. Halperin, Phys. Rev. B 25, 2185 (1982).

\bibitem{MacDonald84prb} A. H. MacDonald and P. Streda, Phys. Rev. B29, 1616
(1984).

\bibitem{Wen92} X. G. Wen, Int. J. Mod. Phys. B 6, 1711 (1992).

\bibitem{Prinz98Science} G. A. Prinz, Science 282, 1660 (1998); S. A. Wolf,
D. D. Awschalom, R. A. Buhrman, J. M. Daughton, S. von Molnar, M. L. Roukes,
A. Y. Chtchelkanova, and D. M. Treger, Science 294, 1488 (2001); D.
Awschalom, D. Loss, and N. Samarth (.eds), Semiconductor Spintronics and
Quantum Computation (Springer, Berlin, 2002).

\bibitem{Dyakonov71} M. I. D'yakonov and V. I. Perel, JETP Lett. 13, 467
(1971); J. E. Hirsch, Phys. Rev. Lett. 83, 1834 (1999); J. Sinova, D.
Culcer, Q. Niu, N. A. Sinitsyn, T. Jungwirth, and A. H. MacDonald, \textit{\
ibid.} 92, 126603 (2004); S. Murakami, N. Nagaosa, and S. C. Zhang, Science
301, 1348 (2003).

\bibitem{Kato04Science} Y. K. Kato, R. C. Myers, A. C. Gossard, and D. D.
Awschalom, Science 306, 1910 (2004); J. Wunderlich, B. Kaestner, J. Sinova,
and T. Jungwirth, Phys. Rev. Lett. 94, 047204 (2005).

\bibitem{Shen04prl} S. Q. Shen, M. Ma, X. C. Xie, and F. C. Zhang, Phys.
Rev. Lett. 92, 256603 (2004); S. Q. Shen, Y. J. Bao, M. Ma, X. C. Xie, and
F. C. Zhang, Phys. Rev. B 71, 155316 (2005).

\bibitem{Schliemann03prb} J. Schliemann, J. C. Egues, and D. Loss, Phys.
Rev. B 67, 085302 (2003); M. G. Pala, M. Governale, Zulicke, and G.
Iannaccone, \textit{ibid.} 71, 115306 (2005).

\bibitem{Rashba60} E. I. Rashba, Fiz. Tverd. Tela (Leningrad) 2, 1224 (1960)
[Sov. Phys. Solid State 2, 1109 (1960)]; J. Lou, H. Munekata, F. F. Fang,
and P. J. Stiles, Phys. Rev. B 38, 10142 (1988).

\bibitem{Nitta97prl} J. Nitta, T. Akazaki, H. Takayanagi, and T. Enoki,
Phys. Rev. Lett. 78, 1335 (1997).

\bibitem{Reynoso04prb} A. Reynoso, G. Usaj, M. J. Sanchez, and C. A.
Balseiro, Phys. Rev. B 70, 235344 (2004);

\bibitem{Dimitrova04xxx} A similar commutator was considered by O. V.
Dimitrova, cond-mat/0407612, and by E. I. Rashba, Phys. Rev. B 70, 201309
(2004). In the absence of an external magnetic field, they observed that $
j_{s}^{z}=-(i\hbar /2m\lambda )[V_{R},\sigma _{x}]$.

\bibitem{Shen04prb} S. Q. Shen, Phys. Rev. B 70, 081311 (2004).

\bibitem{Thouless93prl} D. J. Thouless, Phys. Rev. Lett. 71, 1879 (1993); H.
Hirai and S. Komiyama, Phys. Rev. B 49, 14012 (1994).
\end{thebibliography}
\end{document}